# The Stirred Tank Reactor Polymer Electrolyte Membrane Fuel Cell

Jay Benziger, E. Chia, E. Karnas, J. Moxley, C. Teuscher, I.G. Kevrekidis
Department of Chemical Engineering
Princeton University
Princeton, NJ 08544

**Abstract**

The design and operation of a differential Polymer Electrolyte Membrane (PEM) fuel cell is described. The fuel cell design is based on coupled Stirred Tank Reactors (STR); the gas phase in each reactor compartment was well mixed. The characteristic times for reactant flow, gas phase diffusion and reaction were chosen so that the gas compositions at both the anode and cathode are uniform. The STR PEM fuel cell is one-dimensional; the only spatial gradients are transverse to the membrane. The STR PEM fuel cell was employed to examine fuel cell start-up, and its dynamic responses to changes in load, temperature and reactant flow rates. Multiple time scales in systems response are found to correspond to water absorption by the membrane, water transport through the membrane and stress-related mechanical changes of the membrane.

## I. INTRODUCTION

Fuel cells are multiphase chemical reactors where two sequential chemical reactions are coupled by transport of the intermediate products between catalysts. The reactants are fed to two sides of the reactor, separated by an ion-conducting barrier. A catalytic reaction occurs on one side of the barrier producing an intermediate product that is transported across the barrier to a second catalyst where it reacts with the second reactant to make the final product. A simplified version of the Polymer Electrolyte Membrane (PEM) Fuel Cell is shown in Figure 1.(Bokris and Srinivasan 1969; Blomen and Mugerwa 1993; EG&G Services 2000; Costamagna and Srinivasan 2001) Hydrogen molecules are fed to the anode side of a cation conducting polymer membrane in contact with a catalyst. The hydrogen molecules react on the anode catalyst producing the intermediate product – protons and electrons. The protons are transported across the PEM and the electrons pass through an external circuit where they encounter oxygen molecules on the cathode side of the membrane. The protons, electrons and oxygen react on the cathode catalyst surface and make water.

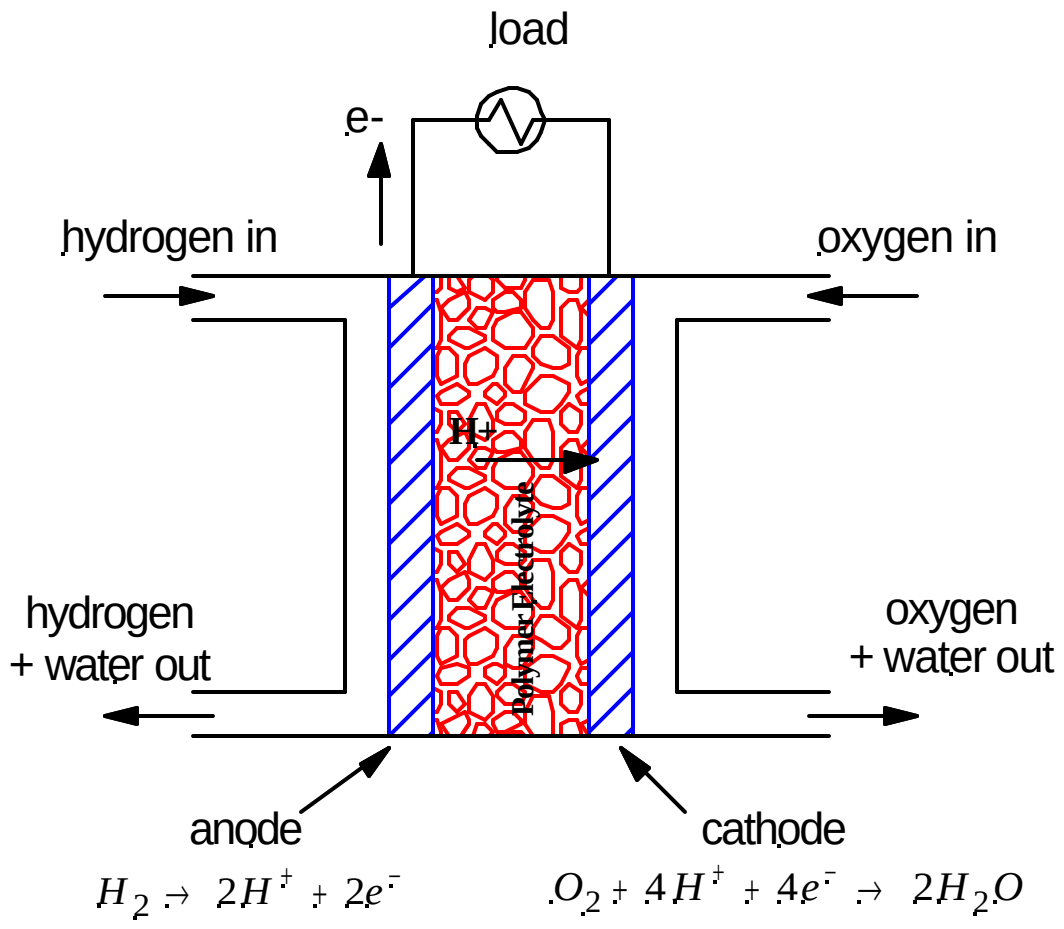

Figure 1. Hydrogen-oxygen PEM fuel cell. Hydrogen molecules dissociatively adsorb at the anode and are oxidized to protons. Electrons travel through an external load resistance. Protons diffuse through the PEM under an electrochemical gradient to the cathode. Oxygen molecules adsorb at the cathode, are reduced and react with the protons to produce water. The product water is absorbed into the PEM, or evaporates into the gas streams at the anode and cathode.

We introduce here a reaction engineering approach to analyze PEM fuel cells. Chemical Reaction Engineering is the branch of engineering science that analyzes reaction and transport processes; this is done on scales ranging from the macroscopic process scale to the molecular scale. Engineers employ various chemical reactor designs to study reaction and transport phenomena in reacting systems. Stirred tank reactors are operated as differential reactors to obtain reaction kinetics. Plug flow reactors provide information about the coupling of reaction kinetics with convective transport. Two-dimensional reactors permit the engineer to obtain information about diffusion of heat and mass in reacting systems. These model reactors are not optimal for reactant conversion, but are specifically designed to measure system parameters, including effective kinetic and transport properties.(Froment and Bischoff 1979; Levenspiel 1996; Folger 1999)

The design engineer relates reactor performance to the system parameters, using the best available correlations between the *system parameters* and the *system variables*. The reaction rate is correlated with the reactant concentrations, catalyst loading, temperature, pressure, etc. Heat and mass transfer are correlated with temperature, pressure, flow rate, composition, geometry, etc. These mathematical correlations between the system's parameters and variables are the key elements of reaction engineering. Although a molecular-level knowledge of the reaction and transport processes helps develop new catalysts and build mathematical correlations the molecular details are not essential to effectively model a reactor's performance. The mathematical correlations provide the correct level of information to assess the performance of different reactor configurations and design control systems. A common joke is that chemical engineers design chemical reactors without ever knowing what the chemical reaction is!

In all humor there lies some truth, and this leads us to a different approach from previous efforts to measure and model fuel cell performance. We are seeking a *prescriptive* model of the PEM fuel cell, which describes the system response as a function of the parameters that the operator can control. We employ an experimental approach based on experience with heterogeneous catalytic reactors; reactor performance for ammonia synthesis, petroleum cracking and reforming, nitric acid synthesis and many other chemical processes have been very successfully modeled knowing only empirical correlations for rate expressions and transport processes.(Froment and Bischoff 1979) Experimental results from simplified experimental reactors were employed to develop mathematical descriptions of the system variables as functions of the system parameters that fit the data over the relevant range of operating conditions.

The mechanistic details of the elementary reaction steps and transport processes in the porous catalyst are not necessary to predict the reactor performance. We believe an analogous approach can be beneficial to predicting fuel cell performance. (We are not dismissing the importance of molecular mechanisms; they are critical in guiding the development of new catalysts and membranes to improve the fuel cells!)

There are a number of excellent models of the transport processes and the detailed chemical reactions at the electrocatalyst surfaces in PEM fuel cells.(Bernardi 1990; Springer, Zawodzinski et al. 1991; Bernardi and Verbrugge 1992; Baschuk and Li 2000; Dutta, Shimpalee et al. 2000; Thampan, Malhotra et al. 2000; Natarajan and Van Nguyen 2001; Springer, Rockward et al. 2001; Van Nguyen and Knobbe 2003) These models have included the molecular details of electron transfer reactions at electrode surfaces, transport of the reactants and products through multiple layers associated with the electrodes, as well as the transport of water and protons through the polymer electrolyte. Steady-state current/voltage response characteristics of a PEM fuel cell have been fit by these models. However, these models are complex and they have not validated with *dynamic* behavior of PEM fuel cells. Indeed, these models have emphasized steady state performance.

We designed and constructed an "ideal" experimental fuel cell to examine fuel cell dynamics. The model fuel cell is a one-dimensional differential reactor with uniform, independently controllable, well-defined gas phase compositions at the anode and cathode. The model fuel cell can be thought of as two stirred tank reactors coupled by a membrane; we refer to this model fuel cell as the STR PEM fuel cell. The one-dimensional structure of the STR PEM fuel cell greatly simplifies the dynamic response to changes in system parameters. Furthermore, the performance of the STR PEM fuel cell can be used as the basis to scale and predict the performance of larger more complex fuel cell reactor systems.

This paper will describe the STR PEM fuel cell and the rationale behind its design. The STR PEM fuel cell is compared to existing fuel cell test stations to illustrate its unique capabilities. A system's analysis is presented to identify the key control parameters affecting the operation of PEM fuel cells. A reaction engineering model of a differential element in a PEM fuel cell is presented to show the essential information required to predict the dynamic behavior. Finally, we present results of the start-up of PEM fuel cells and the responses of the fuel cell to changes in load, temperature and reactant flow.

## II. THE DIFFERENTIAL PEM REACTOR

### *II.1. The STR PEM Design*

A schematic of our STR PEM fuel cell is shown in Figure 2. The membrane-electrode-assembly (MEA) was pressed between two machined graphite plates and sealed with a silicon rubber gasket. Gas plenums of volume, V ~0.2 $cm^3$, were machined in graphite plates above a membrane area of ~1 $cm^2$. There were several pillars matched between the two plates to apply uniform pressure to the MEA. Hydrogen and oxygen were supplied from commercial cylinders (Airco) through mass flow controllers at flow rates, Q~1-10 $cm^3$/min (mL/min). The residence times of the reactants in the gas plenums (V/Q) were greater than the characteristic diffusion time ($V^{2/3}/D$), ensuring uniformity of the gas compositions. The cell temperature was controlled by placing the graphite plates between aluminum plates fitted with cartridge heaters connected to a temperature controller. The entire fuel cell assembly was mounted inside an aluminum box to maintain better temperature uniformity (see Figure 2B).

Gas pressure was maintained in the cell by placing spring loaded pressure relief valves (Swagelok) at the outlets. Tees were placed in the outlet lines (inside the aluminum box) with relative humidity sensors in the dead legs of the tees. The water content of the outlet streams was measured with humidity sensors (Honeywell HIH 3610), and the temperature at the humidity sensor was measured with a thermocouple in the gas line. The relative humidity sensors had to be sufficiently heated to avoid liquid condensation on the capacitive sensing element, but they also had to be kept below 85° to protect the amplifier circuit on the sensor chip.

Any suitable membrane-electrode-assembly (MEA) can be tested in the STR PEM fuel cell. We report here results using an MEA consisting of a Nafion™ 115 membrane pressed between 2 E-tek electrodes (these consist of a carbon cloth coated on one side with a Pt/C catalyst). The catalyst weight loading was 0.4 mg-Pt/$cm^2$. The electrodes were brushed with solubilized Nafion solution to a loading of ~4 mg-Nafion/$cm^2$ before placing the membrane between them. (Raistrick 1989) The assembly was hot pressed at 130°C and 10 MPa. Copper foils were pressed against the graphite plates and copper wires were attached to connect to the external load resistor.

The current and voltage across the load resistor were measured as the load resistance was varied. A 10-turn 0-20 Ω potentiometer was connected in series with a 10-turn 0-500 Ω potentiometer. The load resistance was varied from 0-20 Ω to obtain a polarization curve (IV). To examine the low current range of the polarization curve the resistance would be increased over the range of 0-500 Ω. The voltage across the load resistor was read directly by a DAQ board. The current through the load resistor was passed through a 0.2 Ω sensing resistor and the differential voltage across the sensing resistor was amplified by a factor of 100 with an Analog Devices AMP02 Instrumentation Amplifier and read by the DAQ board. An IV curve was typically collected and stored in ~ 100s.

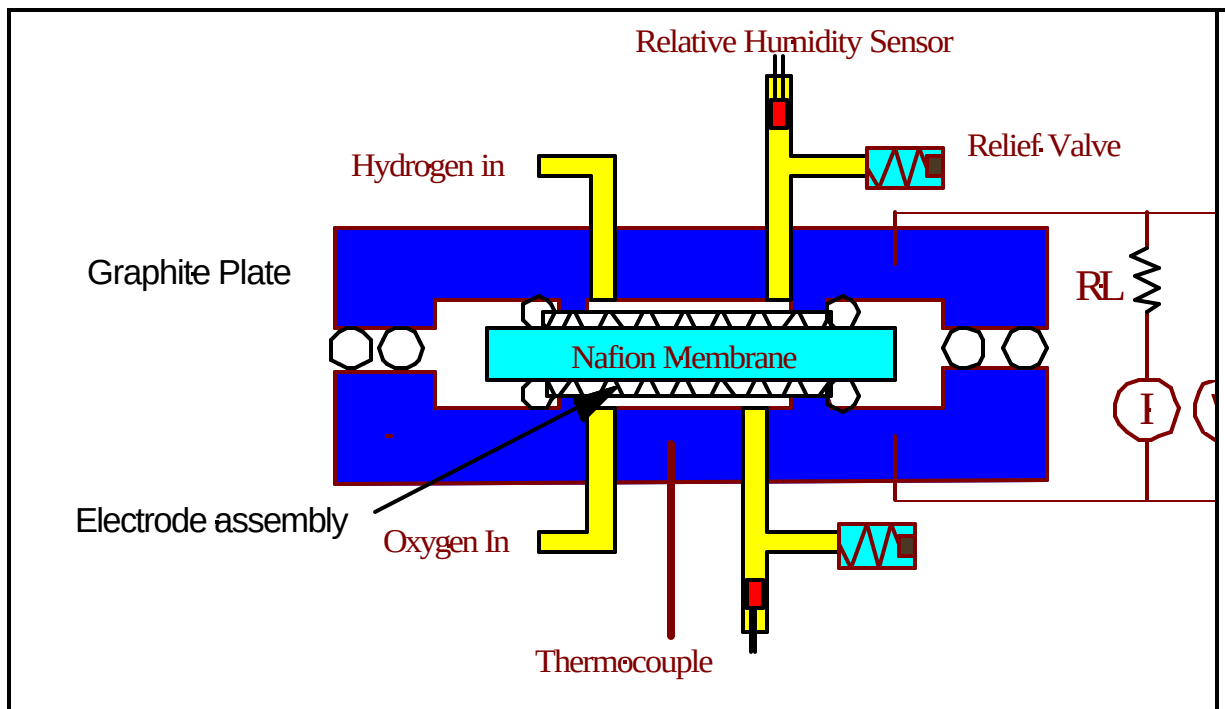

Figure 2A. Schematic of the STR PEM fuel cell. The exposed electrode area was ~1.5 $cm^2$ on each side, with gas volumes above the anode and cathode of ~0.2 $cm^3$. The MEA employed Etek electrodes and a Nafion 115 membrane.

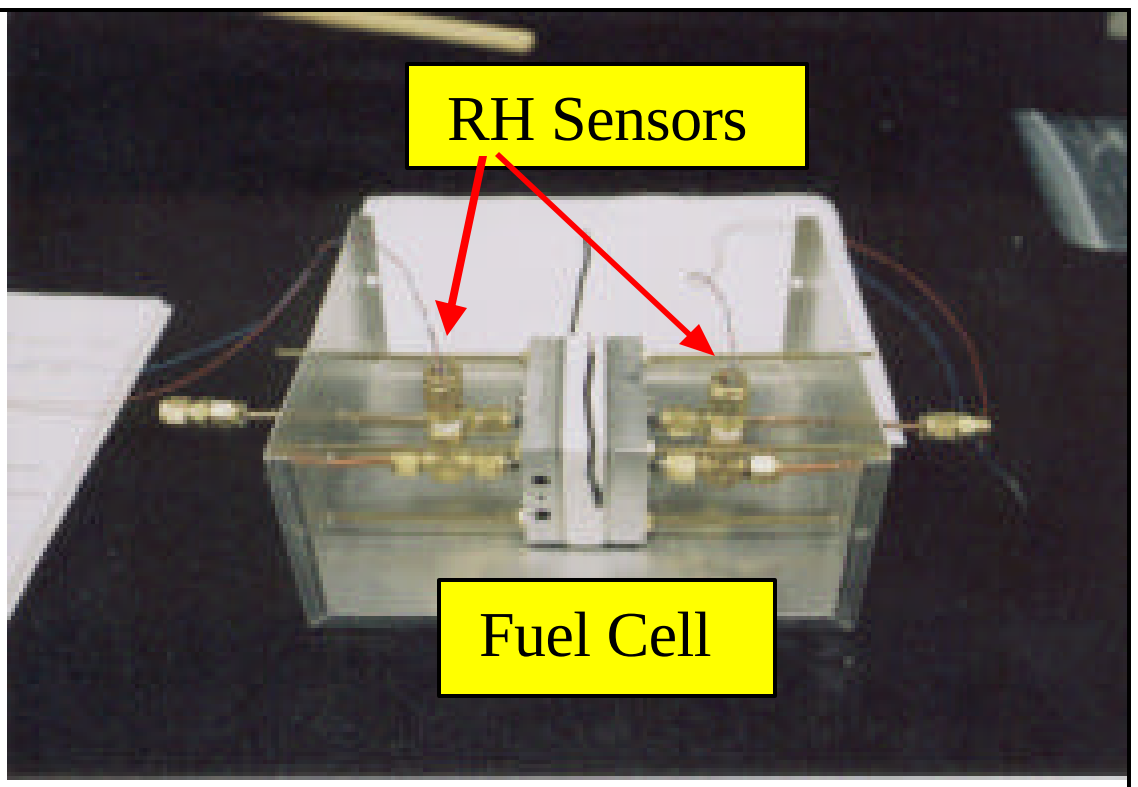

Figure 2B. Photograph of the STR PEM fuel cell. The graphite plates were fitted into Teflon plates and sandwiched between heated Al blocks. Relative humidity sensors measured the temperature and RH in the effluent streams.

*II.2. Comparison of the STR PEM and Serpentine Flow PEM Test Stations*

The fundamental difference between our STR PEM fuel cell reactor and the standard PEM fuel cell test station is associated with the gas flow fields. Figure 3 compares the serpentine flow fields for a GlobeTech Fuel Cell Test Station 2 and our STR PEM fuel cell reactor. The GlobeTech test station has MEA area of 5 $cm^2$, with serpentine flow channels approximately 100 mm long and 1 $mm^2$ cross-sectional area. In the STR PEM the MEA area is ~ 1 $cm^2$, flow channels are approximately 14 mm long with a cross-sectional area of 4 $mm^2$. Mixing in the gas flow channels is characterized by the ratio of diffusive transport (D/L =diffusivity/length of flow channel) to convective transport (u=gas velocity). When $D/uL>1$ diffusive mixing dominates over convective flow and there will be homogeneity in the fluid composition.

The characteristic dispersion number at both the anode and cathode is =1 when the feed flow rates to the STR PEM <10 $cm^3$/min (corresponding to a current density of 1.4 A/$cm^2$ at 100% hydrogen utilization). In contrast, the dispersion number is <0.02 for the serpentine flow channels with flow rates of 50 $cm^3$/min (also corresponding to an average current density of 1.4 A/$cm^2$ at 100% hydrogen utilization). Diffusive mixing in the STR PEM homogenizes the gas phase composition at each electrode. However, convection in the serpentine flow PEM fuel cell test station results in compositional variations along the length of the flow channels. In chemical reaction engineering jargon, the Globe Tech test station is a *plug flow reactor*, whereas our differential reactor is a *stirred tank reactor*. The gas phase uniformity above the anode and cathode simplifies the analysis of the STR PEM data. The system is one dimensional; only gradients transverse to the membrane are important. The current density, or reaction rate, in the STR PEM fuel cell is also spatially uniform; at steady state the reaction rate is equal to the difference between the molar flows of the feed and effluent.

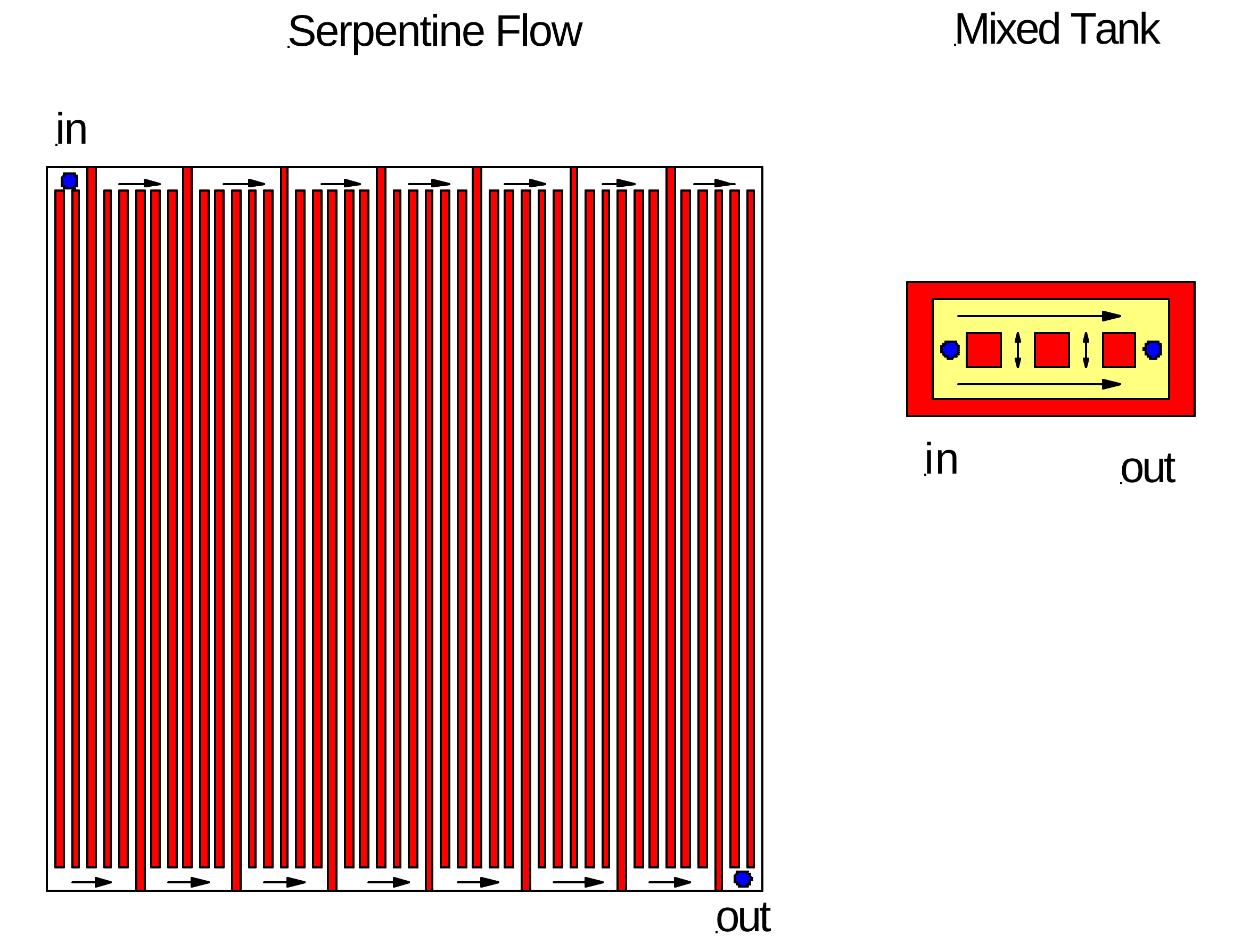

Figure 3. The flow fields, machined in graphite plates, for a "typical" (GlobeTech) PEM test station and for the STR PEM reactor. The Serpentine flow channels cover an area of 5 cm$^2$, while the STR PEM fuel cell covers an area of ~1 cm$^2$. The two different configurations are drawn to the same scale. The open plenum area of the STR PEM reactor permits sufficient diffusive mixing to give near uniform gas phase composition.

It is possible to operate the serpentine flow channel test station in a differential mode by limiting the reactant conversion so the concentration gradients along the flow channel the flow rates are small. Keeping the fractional conversion of the reactants <5% will give nearly homogeneous compositions at the anode and cathode. However, the current density should be limited to 60 mA/cm$^2$ for a serpentine flow channel test station to be differential.

*II.3. System Analysis of PEM Fuel Cells*

The greatest utility of the STR PEM fuel cell reactor is to isolate the dynamics of PEM fuel cells. Specific questions we intend to explore are:

1. How long does it take a PEM fuel cell to start up from different initial conditions? How do the system parameters affect start-up of the fuel cell?
2. How does the PEM fuel cell respond dynamically to changes in load? Temperature? Gas flow rate?
3. How should the system parameters be controlled under conditions of variable load, such as encountered in automotive applications?

The preliminary results presented in this paper illustrate some of the complexities associated with PEM fuel cell dynamics we have identified with our STR PEM fuel cell. Detailed studies and analyses of the fuel cell dynamics will be the subject of future papers.

A real PEM fuel cell reactor is complex. Electrode reactions and transport through the gas channels, diffusion through the electrode layers and transport across the membrane are all coupled. Datta and co-workers have described of the structure of PEM fuel cells and the molecular details of the transport and reaction in the PEM fuel cell.(Thampan, Malhotra et al. 2001) However, these models assume descriptions about transport processes and electrode kinetics and call for data about system variables that cannot be directly measured or controlled. We have followed an engineering approach and considered to what level of detail the system variables in the fuel cell can be described as functions of the parameters under operator control. Our ultimate objective is to develop a good reactor model that includes the essential physics but eliminates unnecessary detail.

Table I summarizes the system variables and system parameters for a PEM fuel cell. The system parameters are under operator control, whereas the system variables describe the local state of the PEM fuel cell. For example, the feed to the fuel cell can be regulated, but the local composition and flow rate along the flow channel are determined by dynamic mass balances. Similarly, water supplied in the feeds is a controlled parameter, while the local membrane water content is a system variable that depends on the balance between water supplied in the feed, water produced at the cathode and water removed in the effluents. The fuel cell current and voltage are *system variables* determined by the state of the membrane and the entire circuit including the controllable external load resistance. We will report operation of the STR PEM reactor under conditions of defined load resistance.

Table I
System Variables and System Parameters for a PEM Fuel Cell

| ***System Variables*** | ***System Parameters*** |
|---|---|
| Reactant flow rates | Reactant feed flow rates |
| Reactant composition | Reactant feed composition |
| Gas Relative Humidity | Heat Input |
| Cell Temperature | External Load Resistance |
| Cell Voltage | Electrode Composition and Structure |
| Cell Current | Membrane Material |
| Membrane Water Content/Resistance | Cell Construction |

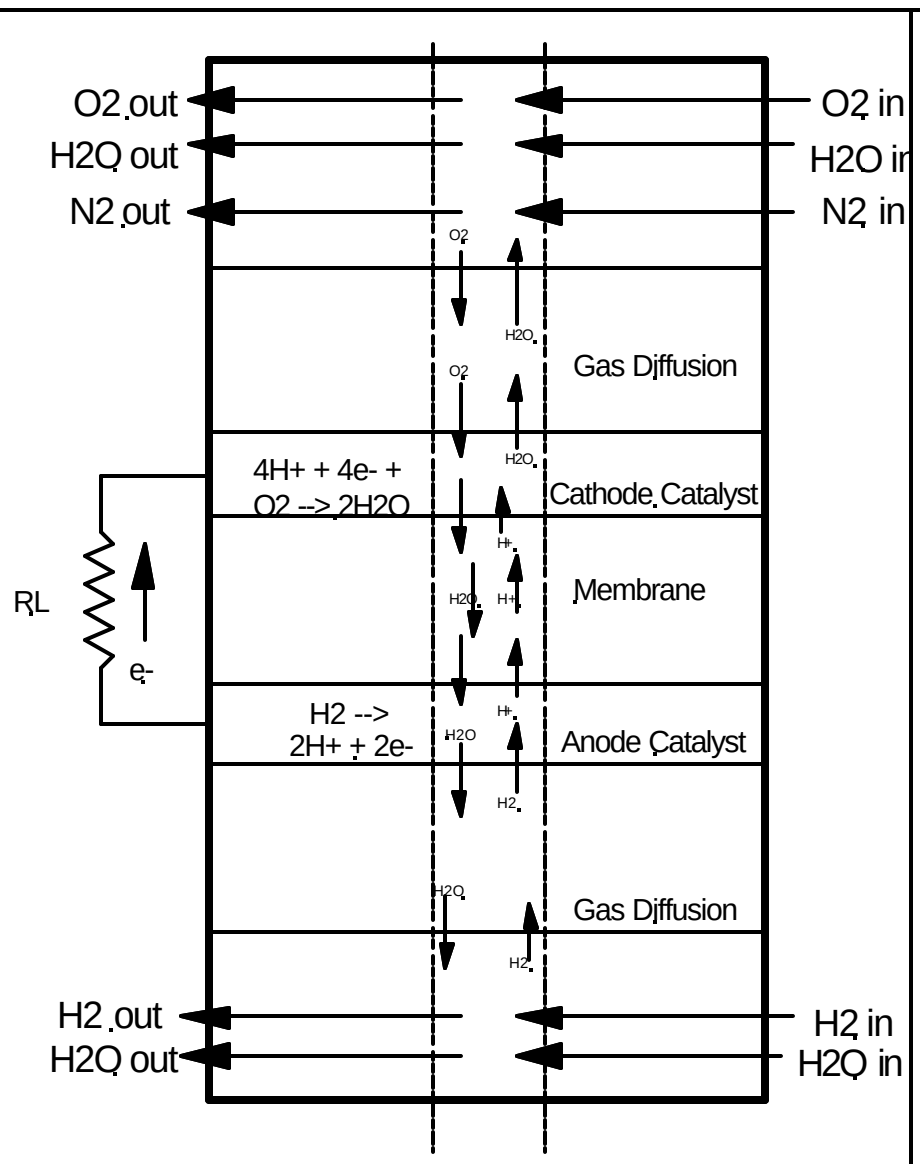

$$\frac{V_A}{RT}\frac{\partial P_{H_2}^{out}}{\partial t} = \left(\frac{Q_A^{in} P_{H_2}^{in}}{RT} - \frac{Q_A^{out} P_{H_2}^{out}}{RT}\right) - \frac{i}{2F} \qquad (1)$$

$$\frac{V_C}{RT}\frac{\partial P_{O_2}^{out}}{\partial t} = \left(\frac{Q_C^{in} P_{O_2}^{in}}{RT} - \frac{Q_C^{out} P_{O_2}^{out}}{RT}\right) - \frac{i}{4F} \qquad (2)$$

$$\frac{\partial N_w^m}{\partial t} = \left(\frac{Q_A^{in} P_w^{A,in}}{RT} - \frac{Q_A^{out} P_w^{A,out}}{RT}\right) + \left(\frac{Q_C^{in} P_w^{C,in}}{RT} - \frac{Q_C^{out} P_w^{C,out}}{RT}\right) + \frac{i}{2F} \qquad (3)$$

$$N_w^m = \frac{r_m A_m (\Delta z)}{EW} l_w \qquad (4)$$

$$l_w = f(a_w^m, T) \qquad ($$

$$a_w^m = f(a_w^A, a_w^C) \qquad (6)$$

$$i = \frac{V_{eff}}{R_L + R_m} \qquad (7)$$

$$R_m = f(a_w^m, T) \qquad (8)$$

$$V_{eff} = \frac{\Delta G^o}{4F} - \frac{RT}{4F}\ln\left(\frac{P_{H_2}^2 P_{O_2}}{P_{H_2O}^2}\right) - h(a_w^m, R_L) \qquad (9)$$

Figure 4A. A schematic of the reaction and diffusion processes in a differential element in a PEM fuel cell. The detailed transport processes through the multiple layers of the membrane-electrode-assembly are illustrated. The flows entering the differential element from the right and the load resistance on the left are the controllable system parameters.

Figure 4B. The model equations shown to the right use a lumped parameter scheme in terms of controllable parameters (inlet flow rates ($Q_i$), inlet compositions ($P_i$) and the external load resistance ($R_L$)) and observable system variables (voltage, current, local composition, membrane water activity).

We can divide the system parameters listed in Table I into two groups. One group of parameters is fixed by the choice of reactor construction, and those remain fixed unless one builds a new reactor. These parameters include choice of membrane, catalyst and flow field. The second group is the parameters that can be manipulated externally during the operation of the fuel cell reactor. These are the feed flow rates, the feed compositions, the heat input (or removal) and the external load resistance. Ideally one would like to know the values of all the system variables during the fuel cell operation. In practice only a few of these quantities are directly observable (measurable). We have designed our STR PEM fuel cell so that the temperature, pressure, gas phase water content in the anode and cathode effluents and the cell current and voltage can be measured. The STR PEM fuel cell minimizes lateral spatial variations, so the current density and gas composition are uniform across the gas-electrode-membrane interface.

*The STR PEM fuel cell can be thought of as a differential element along the flow channel in a PEM fuel cell.* The differential element is shown in Figure 4, along with the differential material balances (we will treat the fuel cell as isothermal in this paper – energy balances will be the topic of a subsequent paper). We employ a lumped parameter model, which emphasizes the functional description of the fuel cell based on controllable parameters and observable variables. Equations 1-9 summarize the model equations in the differential reactor element. Equations 1-3 are mass balances for hydrogen, oxygen and water. Equation 7 represents the reaction rate for water formation, which is equal to ½ the proton current. The remaining equations are the relations between different system variables. The terms in the equations are defined in the nomenclature section at the end of the paper.

*II.4 Reactor Model of a PEM Fuel Cell*

The system parameters for the fuel cell are slightly different from those of a typical chemical reactor. In addition to feed flow rates, composition and temperature, the external load resistance is a new parameter. The fuel cell can be thought of as a set of reactors connected through a set of flow regulators, as shown in Figure 5. Hydrogen molecules are oxidized to protons and electrons, at the anode. The resistances in the membrane and external load regulate the current in the fuel cell (i.e. the flow of protons and electrons. The protons and electrons meet up at the cathode along with the oxygen to produce water. The external load resistance is analogous to a valve that regulates the flow of product out of the anode reactor to the cathode reactor.

The coupling of reactor elements shown in Figure 5 is the basis for our analysis of the fuel cell as a chemical reactor. The system parameters are the feed flow rates, composition, the heat input and the external load resistance. (Our STR PEM fuel cell is small and generates little heat. It is surrounded with a large heat source/sink creating a uniform temperature so temperature may be treated as a fixed system variable.) *We present data with the external load resistance as the independent parameter. This is different than the traditional electrochemistry approach where PEM fuel cells are operated under galvanostatic or potentiostatic control (constant current or constant voltage).* When the chemical reaction is driven by the imposition of an external electrical driving force, such as with electrolysis of water, the current or voltage is a system parameter that can be independently manipulated. However, in a fuel cell the chemical reaction drives the current through the external load, and the load resistance is the system parameter that can be manipulated. Constant current or voltage requires a feedback controller that adjusts the external resistance to maintain the current of voltage. We wish to understand the *autonomous* operation of the PEM fuel cell; operation of the fuel cell under galvanostatic or potentiostatic control distorts the autonomous dynamics and obscures the direct determination of kinetics.

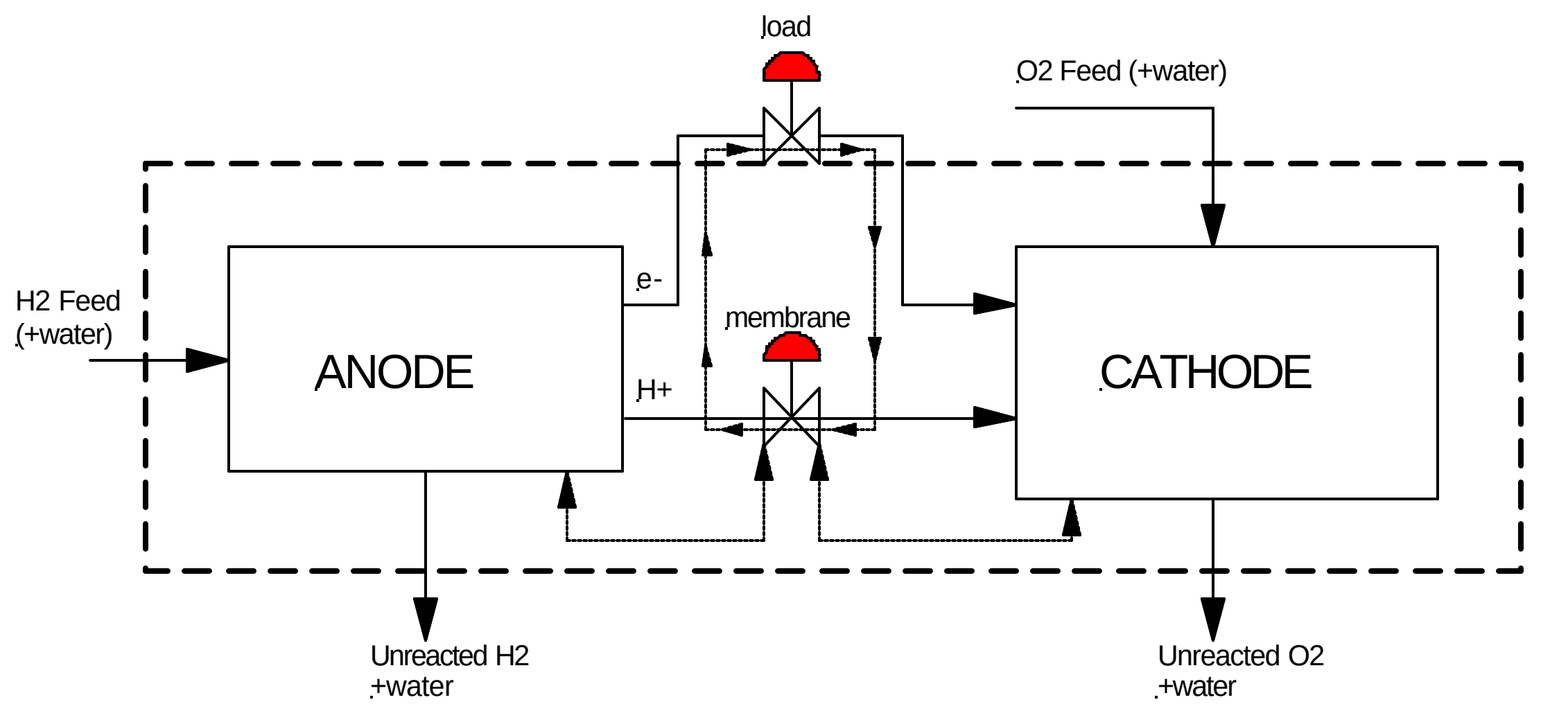

Figure 5. Conceptual reactor coupling in a fuel cell. The heavy dashed line represents the physical boundary of the fuel cell. The feed flow and composition at the anode and cathode are system parameters shown as inputs. The effluents leaving the anode and cathode are system variables. The membrane and the external load resistance for the fuel cell are analogous to valves that regulate flow of the intermediate products from the anode to the cathode. The dashed line going through the valves indicates that the resistance to flow of those two regulating valves is in series. The load resistance is shown external to the fuel cell boundary as it is a system parameter. Water is shown moving between the cathode and anode through the membrane. The water flux depends on the concentration gradients and the current (via electro-osmotic drag).

In the STR PEM fuel cell we set the feed conditions, the temperature and the external load resistance. We measure the effluent flow rate and composition, the voltage across the external load and the current through the external load. The key system variable that we cannot measure directly in our setup is the membrane water activity. The membrane water activity determines the proton conductivity of the membrane, which along with the external load resistance controls the current and voltage associated with the fuel cell. Equation 9 relates the effective fuel cell voltage to the system variables and parameters, the first two terms on the right hand side of equation 9 are the thermodynamic potential. The last term is the overpotential, representing a kinetic limitation. We have expressed the overpotential as a function of water activity in the membrane and load resistance. Normally the overpotential is expressed as a function of the current density. However, the load resistance and membrane resistance, which is a function of the water activity, determine the current and in turn the overpotential.

## III. DYNAMIC OPERATION OF THE STR PEM FUEL CELL

A PEM fuel cell must have sufficient water content for the fuel cell to function; but how much water is sufficient? We show two experiments that illustrate the importance of water in the start-up of PEM fuel cells. A series of experiments are then presented where changes in the system parameters alter the balance between water production and removal and change the water activity in the membrane. The membrane is a reservoir for water, and the resistance of the membrane change as the water inventory changes. The coupling the electrical resistance and the water content results in a feedback loop that causes complex dynamics in the PEM fuel cell.

*III.1. Start-up of the Autohumidification STR PEM Fuel Cell*

Autohumidification refers to fuel cell operation with dry feed gases; the water to humidify the membrane is produced by the fuel cell reaction. Shown in Figure 6 is the current response for start-up of the STR PEM fuel cell operated in the autohumidification mode. Prior to start-up of the STR PEM fuel cell, the initial water content in the membrane and the load resistance were fixed. The polymer membrane was dried by flowing dry oxygen through the cathode chamber at ~100 mL/min and dry nitrogen through the anode chamber at ~100 mL/min for ~12 hours at 60ºC. To humidify the membrane, the oxygen flow was shut off, and 10 mL/min nitrogen flow was passed through a water bubbler at room temperature and into the anode chamber. The relative humidity was measured at the outlet of the anode as a function of time to determine the water uptake by the membrane. After hydrating the membrane to the desired level, the nitrogen flow was stopped. Hydrogen flow at 10 mL/min to the anode and oxygen flow at 10 mL/min to the cathode were initiated, and the current through the load resistor (set at 5 Ω) was measured as a function of time. For initial membrane water concentrations of <0.6 $mg/cm^2$ the fuel cell current decayed with time to near zero (the fuel cell current was “extinguished”). When the water concentration in the membrane was ~0.8 $mg/cm^2$ the fuel cell current “ignited”, rising from an initial value of ~16 mA to a final value of 130 mA. The relative humidity in the effluents followed the same trends as the fuel cell current, when the current decayed the relative humidity in both streams approached zero, and when the fuel cell current increased the relative humidity increased. The critical initial water content for sustained operation corresponds to “ignition” of the fuel cell. When the initial membrane water content is greater than the critical level water production is sufficient to sustain the water content in the membrane. At lesser initial water content the resistance to proton current is too great and evaporation of water from the membrane exceeds water production dehydrating the membrane and extinguishes the current.

Figure 6B shows an experiment where the initial water loading in the membrane was the same, but the external load resistance was changed. The flow rates were still set to 10 mL/min for both $H_2$ and $O_2$ and the fuel cell temperature was 60°C. With an external load resistance of 5 Ω the fuel cell current ignited, rising from ~20 mA to a final value of 150 mA. In contrast when the external load resistance was 20 Ω the current was extinguished, starting at ~7 mA and decaying. This result illustrates how the membrane and external load resistances act in series, and either one could limit the ultimate steady state current.

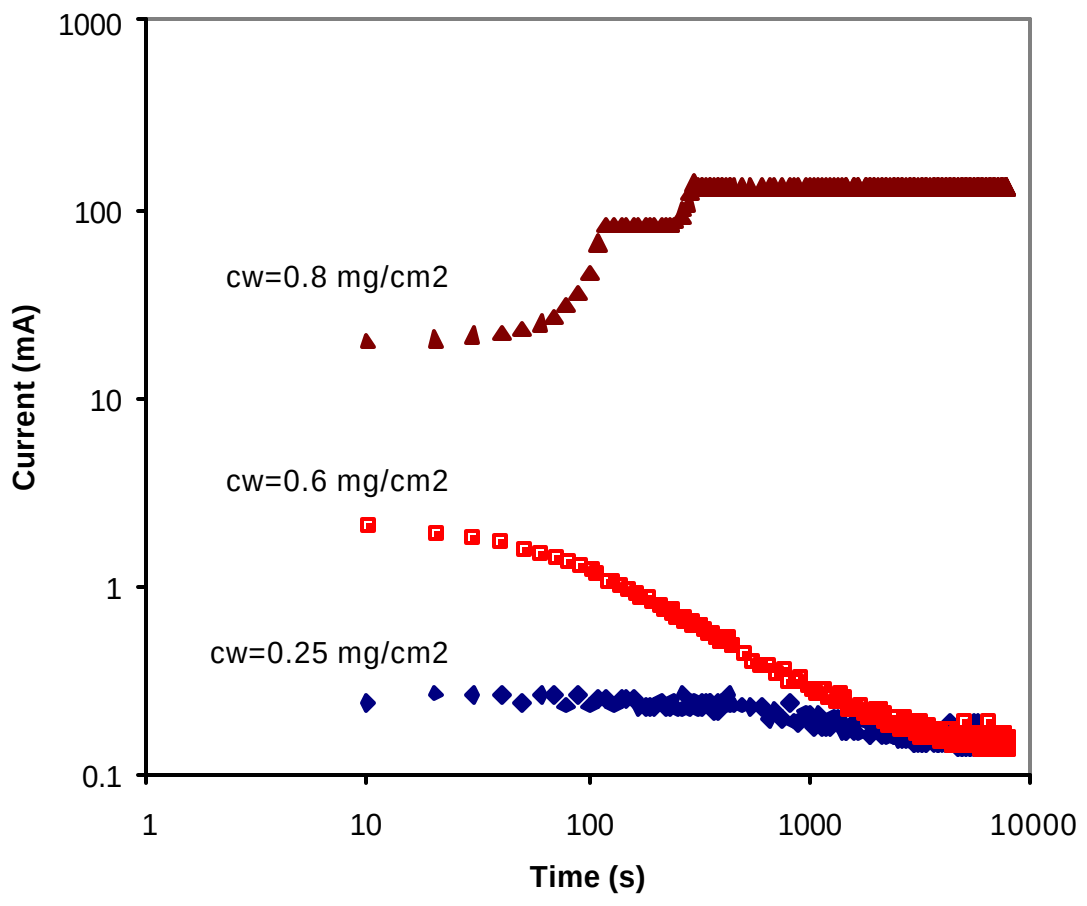

Figure 6A. STR PEM fuel cell start-up from different initial membrane water contents. The membrane was exposed to humidified nitrogen at room temperature to hydrate it, and then heated to 60°C and a flow of 10 mL/min of $H_2$ to the anode and $O_2$ to the cathode was initiated.

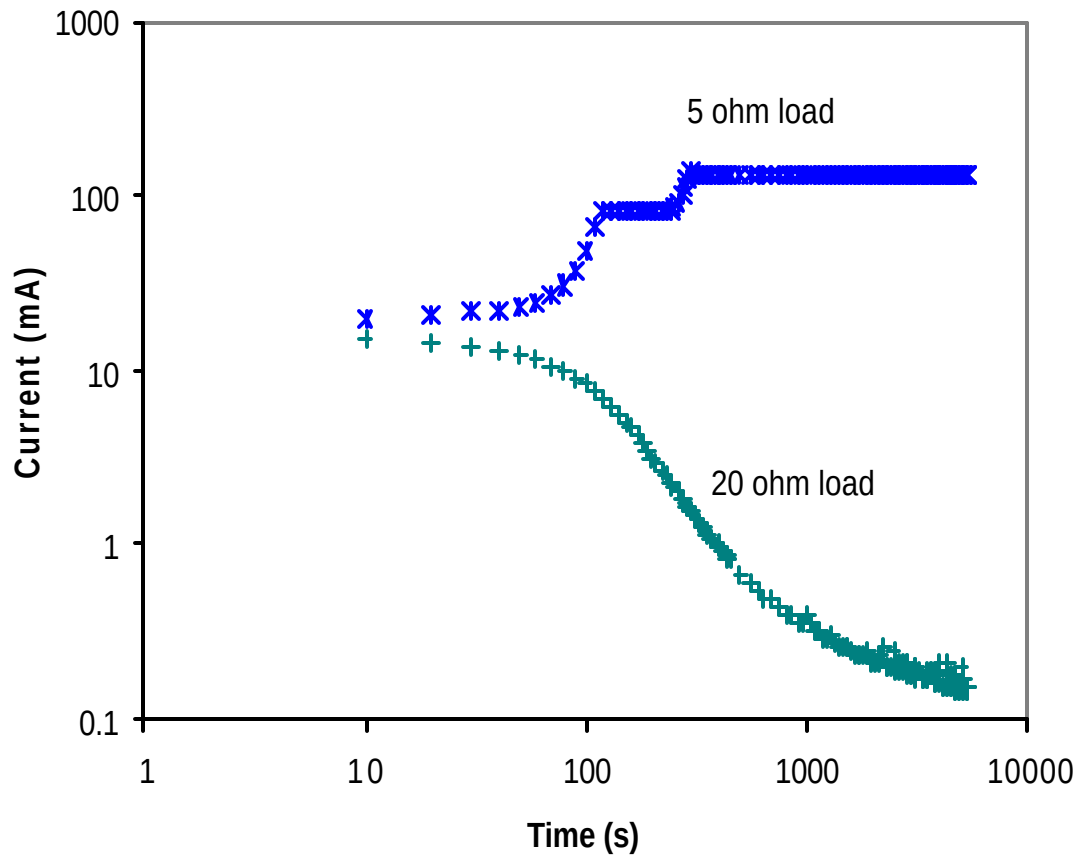

Figure 6B. STR PEM fuel cell start-up with different load resistances. The initial water content in both cases was ~0.8 mg/cm2. The fuel cell was operated at 60°C and a flow of 10 mL/min of $H_2$ to the anode and $O_2$ to the cathode.

*III.2. Critical Humidification of Reactant Feed*

Humidifying the reactant feed may also result in ignition of the fuel cell. Figure 7 shows an experiment where the STR PEM fuel cell was initialized as described above with a "dry" membrane (the membrane was dried by flowing dry gases passing through the fuel cell at 60°C for 12 hours). At time zero dry $O_2$ was introduced to the fuel cell at 10 $cm^3$/min, and 10 cm3/min $H_2$ feed was first passed through a bubbler. The bubbler temperature was controlled with an external heating tape connected to a Variac. Humidification of the anode feed the fuel cell "ignited" as shown in Figure 7A. After ignition the water produced further increases the water activity of the fuel cell effluents. The critical feed humidification for ignition is demonstrated in Figure 7B. Increasing the humidifier temperature from 25 to 30ºC resulted in ignition of the fuel cell current. Further increase in the humidifier temperature to 35ºC resulted in more rapid humidification of the membrane and earlier ignition of the fuel cell, but the final steady state current was the same. The final steady state current depends primarily on the water activity in the membrane, and only indirectly on the water content of the feed (as a threshold for ignition).

The ignition phenomenon results from a positive feedback between the membrane water activity and the reaction rate. Increased membrane water activity decreases the membrane resistance, which according to equation 7 will increase the fuel cell current. The increased current produces more water that will further increase the water activity in the membrane. The current increase is self-limiting. At high membrane water activity liquid water condenses in the catalyst layer of the cathode inhibiting oxygen mass transport to the cathode. This corresponds to a shift in the rate-limiting step in the fuel cell reaction from transport across the membrane to reactant transport from the gas to the cathode surface.

The ignition phenomenon reported here shows a direct analogy to thermal ignition for exothermic reactions in stirred tank reactors. With the autothermal reactor there is a critical initial temperature for ignition. The reactor can also be ignited by preheating the reactor feed.(Liljenroth 1918; van Heerden 1953; Froment and Bischoff 1979; Folger 1999)

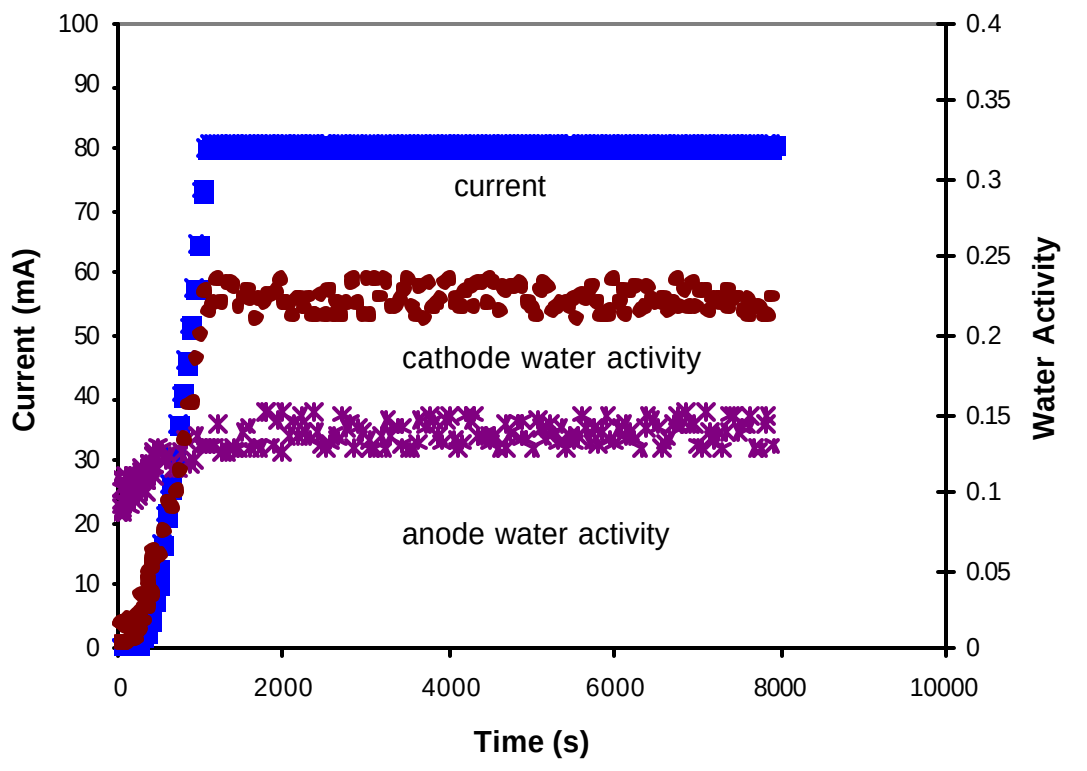

Figure 7A. Start-up of the STR PEM fuel cell operated at 60ºC with feeds of 10 mL/min $O_2$ and 10 mL/min $H_2$. The membrane was initially dry. The hydrogen feed was humidified in a bubbler held at 30ºC. The external load resistance was 7Ω.

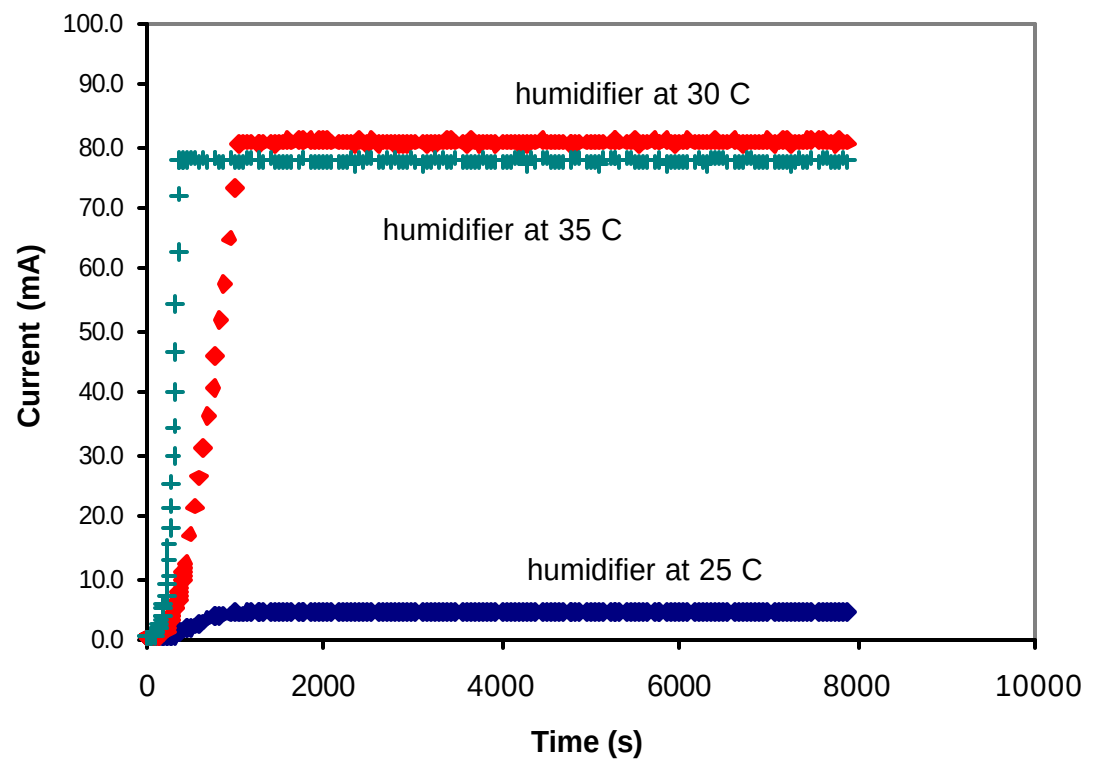

Figure 7B. Start-up of the STR PEM fuel cell operated at 60ºC with feeds of 10 mL/min $O_2$ and 10 mL/min $H_2$. The membrane was initially dry. The hydrogen feed was humidified in a bubbler held at the temperatures shown in the figure. The external resistance was 7Ω.

*III.3. Fuel Cell Response to Changes in Load*

When used for automotive applications, fuel cells need to respond to changes in the load. Changing the load alters the water production changing the balance between water produced and water removed, resulting in a change in the membrane water content. The effect of the load resistance on the water activity can be seen in the polarization curves for the STR PEM fuel cell shown in Figure 8A. The STR PEM fuel cell was operated in the autohumidification mode. The STR PEM fuel cell was equilibrated at 80°C for 12 hr with a fixed load resistance (either 0.2 ? or 20 ?). After equilibration the polarization curve was obtained by sweeping the load resistance between 0.2-20 Ω in 100 s. The relative humidity in the anode and cathode streams changed by <2% while obtaining these polarization curves; for all practical purposes these polarization curves are taken at "constant" membrane water activity.

Figure 8A illustrates that the "instantaneous" polarization curve does not represent a unique characterization of the PEM fuel cell. Operation with different load resistances for extended periods of time resulted in different membrane water activities. The membrane water activity is critical in defining the "instantaneous" polarization curve. The striking feature about Figure 8A is that the two polarization curves cross. Extended operation with a low load resistance produced an MEA with "high" water content, while extended operation with a high load resistance produces an MEA with "low" water content. The MEA with the high water content shows a higher voltage at low currents, indicating a lower activation polarization. At high currents the "high" water content of the MEA shows a lower voltage suggesting the water is limiting mass transport of oxygen to the cathode. The "low" water content MEA has greater activation polarization, but a lower mass transport resistance.

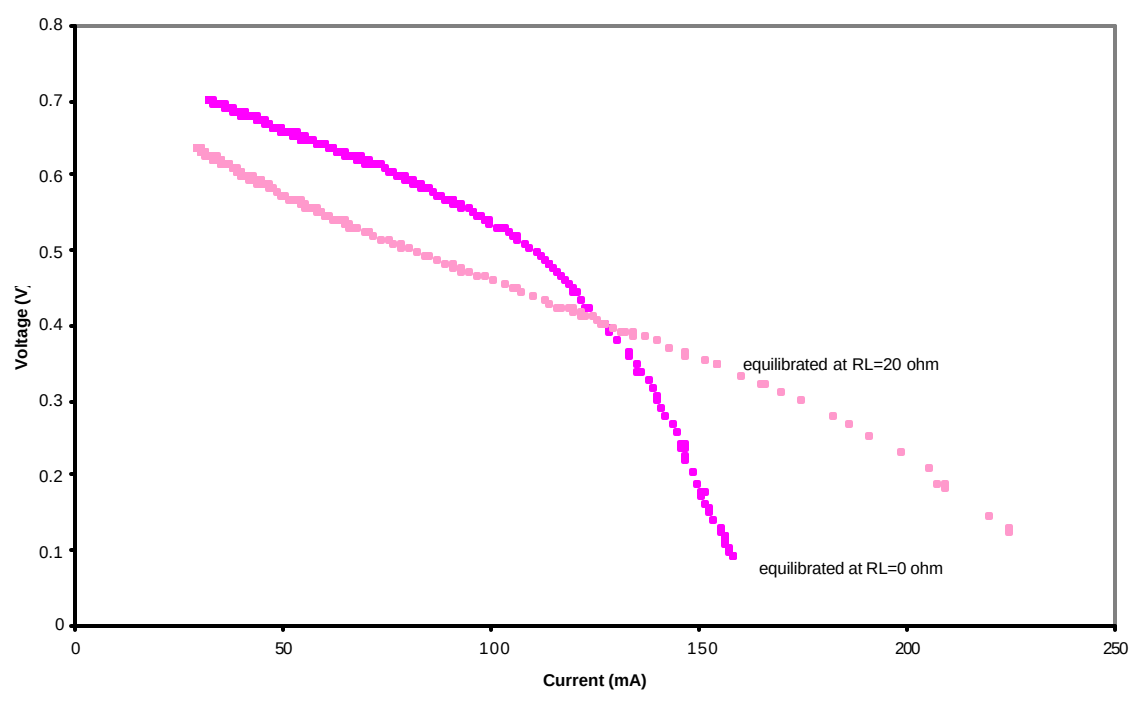

Figure 8A. Polarization curves for STR PEM fuel cell equilibrated with a fixed load resistance at 80°C for 12 hours. The IV curves were recorded by sweeping the load resistance from 0-20 Ω in a period of 100 s.

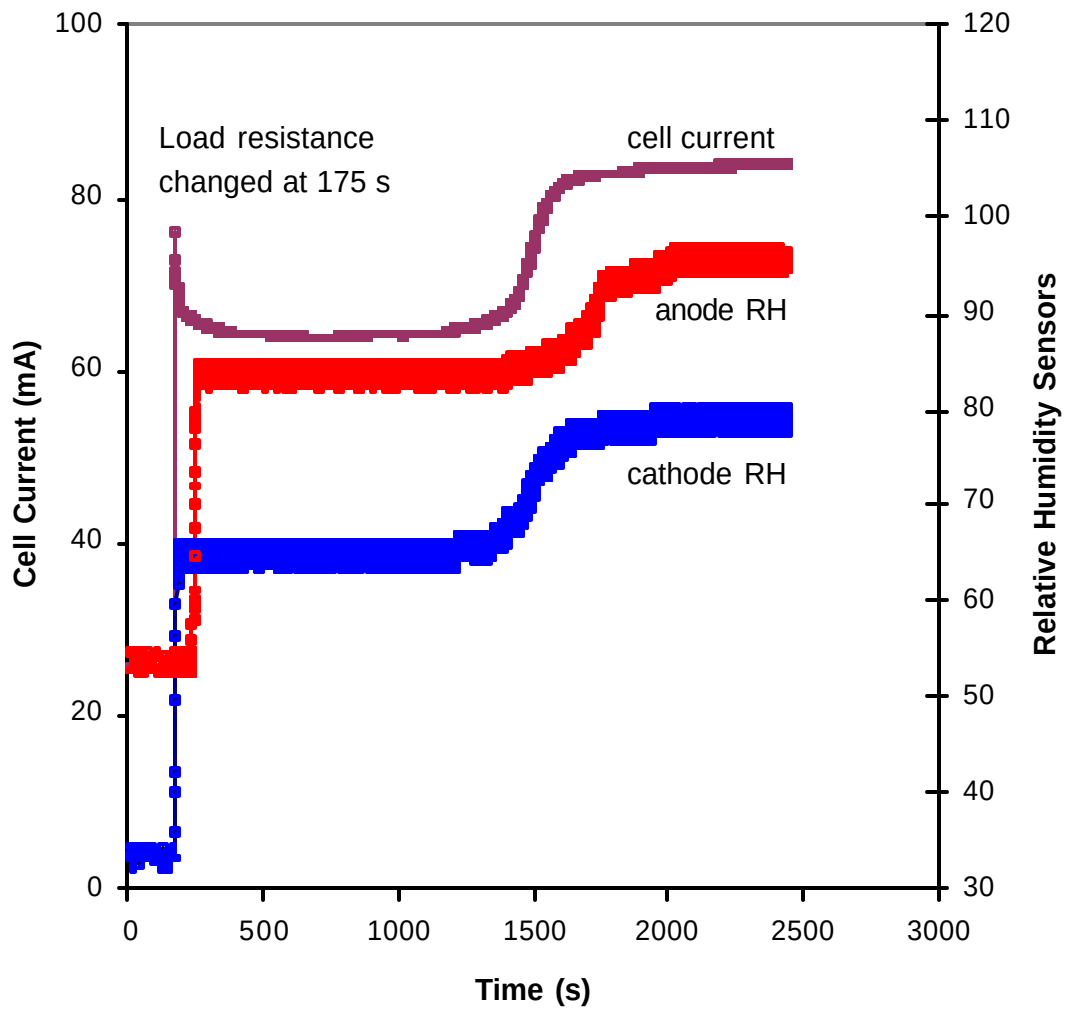

Figure 8B. Dynamic response of the STR PEM fuel cell for switching the load resistance from 20Ω to 7Ω at 80°C. The flow rates were 5 mL/min $H_2$ and 10 mL/min $O_2$. The resistance was switched at 175 s.

The dynamic response of the STR PEM fuel cell to a change in resistive load shows an unusual multi-step process. Figure 8B shows an immediate step response of the current to the change in load, followed by decay to plateau value. There was a subsequent jump in the fuel cell current after 1500 s. The time constant for the rise to the initial plateau was ~10 s. There was a delay of ~100 s in the change of the water activity at the anode relative to the change in current and the change in water activity at the cathode. The jump in current after 1500 s occurred with no changes to any external parameter and was completely unexpected. The cathode relative humidity response tracks the current response; the anode relative humidity response tracked the current but was delayed by ~100 s.

The response times of a PEM fuel cell may surprise many people. It does not fit with the common assumption that PEM fuel cells have response times of milliseconds that make them appropriate for use in automobiles. The data show at least four different time constants associated with the dynamic response of the fuel cell to changes in load. The initial response that occurs almost instantaneously must correspond to the change in current at constant membrane water content. The other time constants must correspond to transport processes, and changes in the membrane properties that result from changes in membrane water content.

What physical processes can account for the PEM fuel cell responses?

We can compare various time constants associated with the PEM fuel cell. Four of the key time constants are listed in Table II. They include: the characteristic reaction time of the PEM fuel cell ($\tau_1$), the time for gas phase transport across the diffusion layer to the membrane electrode interface ($\tau_2$), the characteristic time for water to diffuse across the membrane from the cathode to the anode ($\tau_3$), and the characteristic time for water produced to be absorbed by the membrane ($\tau_4$). Approximate values for the physical parameters have been used to obtain order of magnitude estimates of these time constants. The estimated time constants shown in Table II indicate that the response times of ~100 s are associated with water uptake and transport through the membrane.

The 100 s time for water transport through the membrane is evident in the delay of the response of the water activity in the anode effluent compared to the rise in current. Water absorption by the membrane has a time constant of ~10-100 s. Operating at a current density of 1 A/cm$^2$ it would take ~100 s to saturate a dry membrane with water, assuming no water evaporates into the gas effluents from the fuel cell. Likewise when the load resistance is increased the finite evaporation rate results in a slow decay to steady state. *The membrane acts as a reservoir for water as the external load resistance, reactant flow rates, and temperature changes alter the balance between water production and water removal.*

Table II
Characteristics Times for PEM Fuel Cells

| | Physical significance | | Approximate Value |
|---|---|---|---|
| $\tau_1$ | Characteristic time for reaction rate relative to reactor volume | $t_1 = \frac{V_R}{i} \sim \frac{(0.1\,cm^3/cm^2)}{(1\,A/cm^2)}$ | 0.1 s |
| $\tau_2$ | Characteristic diffusion time across gas diffusion layer | $t_2 = \frac{(\ell_{diffusion\,layer})^2}{(D_{gas}^{eff})} \sim \frac{(0.03cm)^2}{(0.01cm^2/s)}$ | 0.1 s |
| $\tau_3$ | Characteristic diffusion time for water across membrane from cathode to anode | $t_3 = \frac{(\ell_{membrane})^2}{(D_{water}^{membrane})} \sim \frac{(0.01cm)^2}{(10^{-6}\,cm^2/s)}$ | 100 s |
| $\tau_4$ | Characteristic time for water production relative to sulfonic acid density | $t_4 = \frac{l N_{SO3}}{i} \sim \frac{5(2.3\times10^{-5} mol/cm^2)}{(1\,A/cm^2)}$ | 100 s |

The 1500 s time constant for the second jump in the current shown in Figure 8B is still not well understood. We have recently measured the stress relaxation of Nafion. A Nafion 117 sample was strained to 50%, beyond its yield point for plastic deformation, and the time stress was measured as a function of time. At room temperature the stress took ~4000 s to relax to a constant value. We believe the jump in the current after 1500 s is due to the relaxation of the stress in the membrane. Increased membrane water content results in the membrane swelling which puts the membrane under stress. Relaxation of the membrane stress appears to reduce the electrical resistance of the membrane.

*III.4. Response to Temperature Changes*

The dynamic response of the fuel cell to changes in the temperature can be used to explore the dynamic response of fuel cells to changes in heat dissipated. The response of the STR PEM fuel cell to a change in temperature is shown in Figure 9. The fuel cell was operated with fixed feed flow rates of 10 mL/min $H_2$ at the anode and 10 mL/min $O_2$ at the cathode. The load resistance was fixed at 2 Ω. After changing the setpoint on the temperature controller, the temperature, cell current and voltage and the relative humidity in the anode and cathode effluents were monitored. The temperature controller could actively heat the cell from 70 to 90°C in ~200 s; cooling was passive, so it took ~1400 s for the temperature to fall from 90 to 70°C. As the temperature fell, the current and effluent relative humidities all increased. The decrease in the water vapor pressure with temperature reduced the water removal rate. With less water removed the water activity in the membrane increased resulting in a higher current. The relative humidity in the effluents also increased because the vapor pressure of water is lower, so even for the same partial pressure of water in the effluent streams the relative humidity is greater.

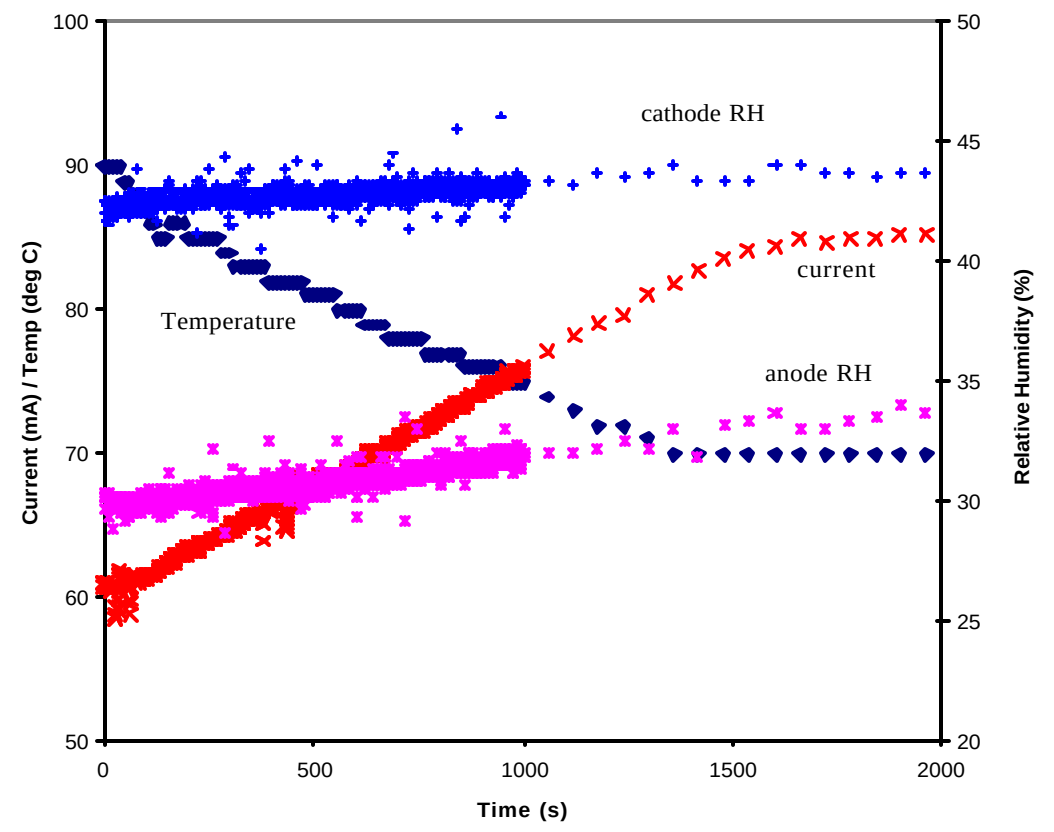

Figure 9A. STR PEM fuel cell response to a decrease in temperature from 90 to 70°C. $H_2$ flow and $O_2$ flow were both 10 mL/min and the external load resistance was 2 Ω.

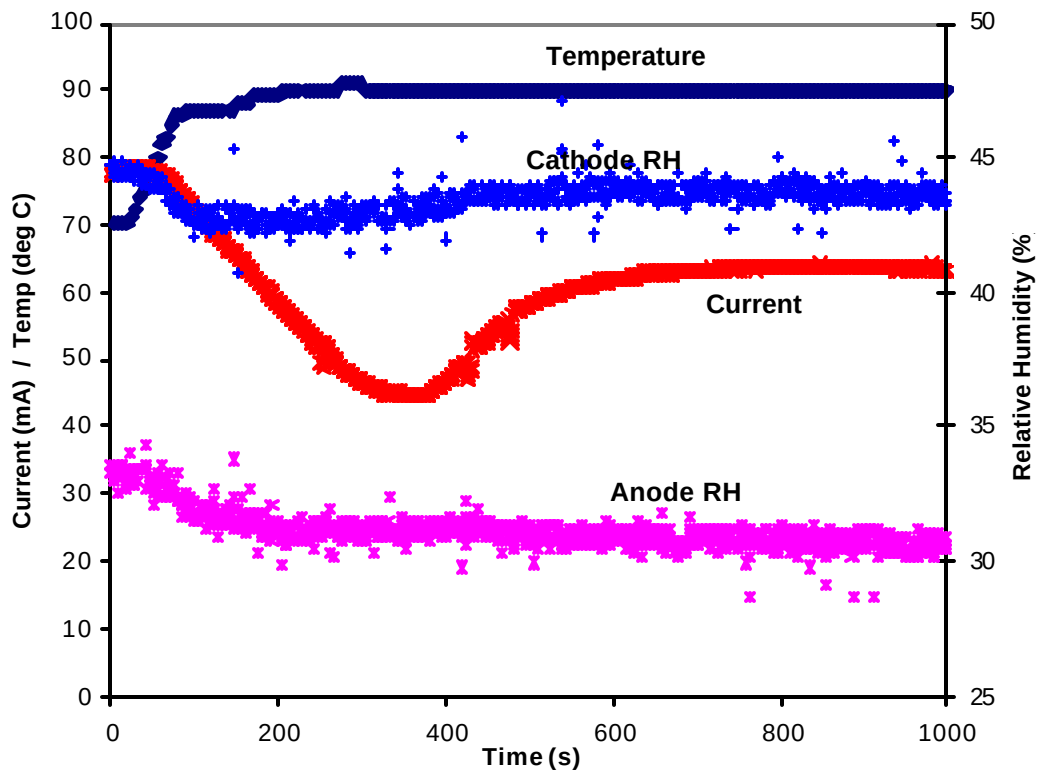

Figure 9B. STR PEM fuel cell response to an increase in temperature from 70 to 90°C. $H_2$ flow and $O_2$ flow were both 10 mL/min and the external load resistance was 2 Ω.

The response of the STR PEM fuel cell to an increase in temperature was surprising. The current and relative humidity in both effluent streams initially decreased. The current and cathode relative humidity went through minima before approaching the steady state. This suggests that evaporation from the MEA is faster initially than the diffusion of water across the membrane in the MEA. It took the current and relative humidity over 700 s to reach steady state, while the temperature was at steady state after only 200s. The long transition to steady state resulted from equilibration between water in the membrane and water at the membrane electrode interfaces. The differences in the relative humidity responses at the anode and cathode are indicative of the complex coupling between water transport into and through the membrane and water production at the cathode.

Raising the temperature increased the water vapor pressure which increases the water removal rate from the fuel cell. At constant water activity the membrane resistance has a weak temperature dependence.(Yang 2003) Increasing the temperature from 80-140°C, decreases the resistivity of Nafion by 50%. The water vapor pressure increased by approximately 700% over the same temperature span. With all else the same, the higher temperature will reduce the water content in the membrane and ultimately reduce the current.

### III.4. DYNAMIC RESPONSE TO CHANGES IN REACTANT FLOW RATES

The flow rates to the anode of the fuel cell should be varied during operation to achieve high hydrogen utilization. The dynamic response of the STR PEM fuel cell to changes in $H_2$ flow is shown in Figure 10. The STR PEM fuel cell was equilibrated for 12 hours with a $H_2$ flow of 1 mL/min and then the $H_2$ flow was rapidly increased to 10 mL/min. The oxygen flow to the cathode was kept constant at 10 mL/min. The cell temperature was 80°C and the external load resistance was 2 Ω. The cell current increased from 3 mA to ~80 mA rapidly during the first 10 s after the change in flow rate. The current increased more slowly over the next 100 s and leveled off at ~100 mA. The relative humidity at the cathode began to increase 10 s after the flow rate was increased and the current increased. The cathode relative humidity continued to climb steadily as the current leveled off. The current jumped suddenly after ~650 s from 105 to 145 mA, and the relative humidity at the cathode increased much more slowly after 650 s. The anode relative humidity showed a steady increase for the entire 2000 s of the test run.

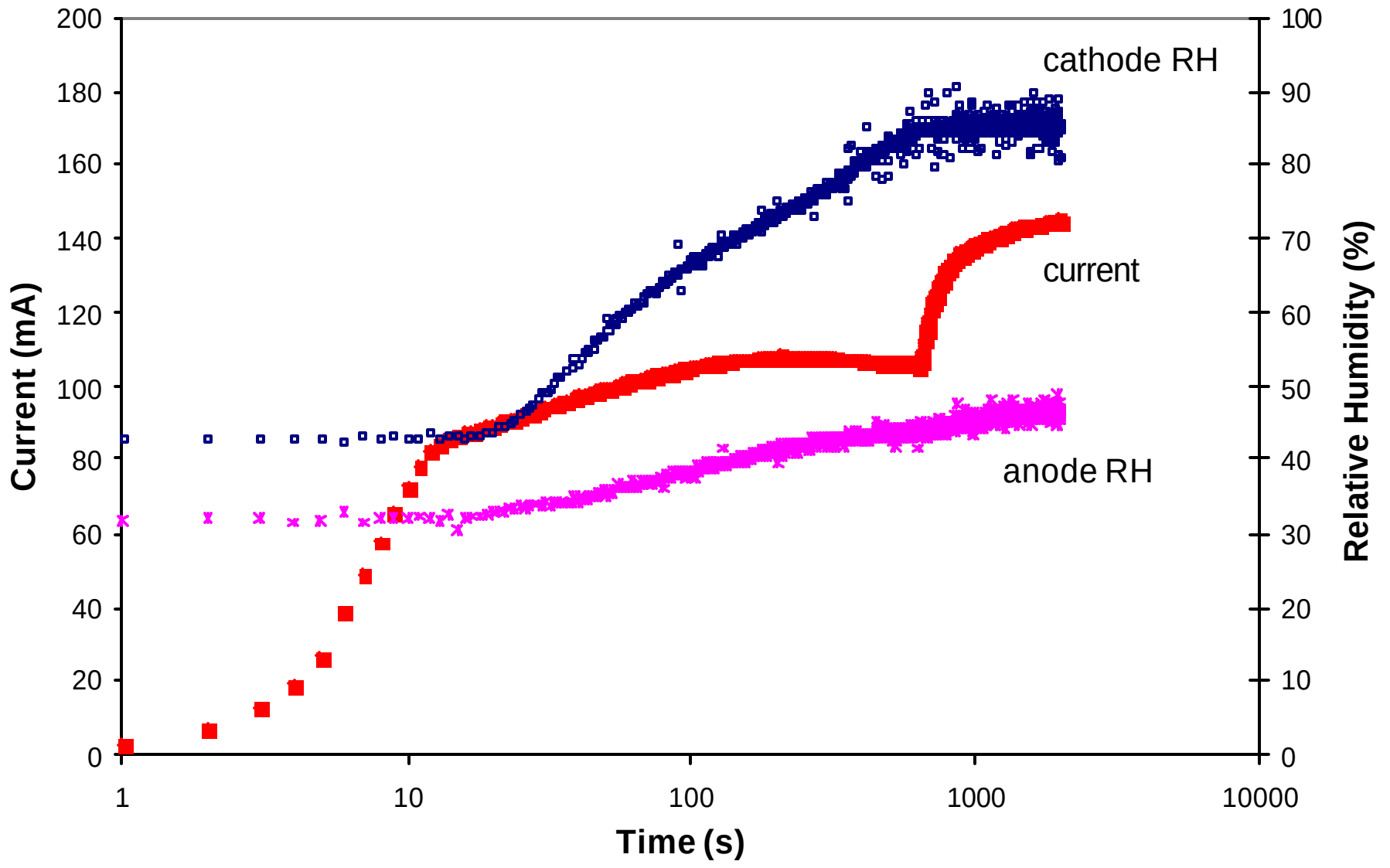

Figure 10. Response of the STR PEM fuel cell to a change in anode flow rate. The $H_2$ feed to the anode was increased from 1 mL/min to 10 mL/min at time 0. The $O_2$ feed to the cathode was constant at 10 mL/min, the cell temperature was 80°C and the load resistance was 2Ω.

The dynamic response during the first 650 s shown in Figure 10 expected. Increasing the supply of hydrogen increased current and water production. With more water produced, more water exited through the effluent streams, because of increased relative humidity. The jump in the current after 650 s is surprising. We believe the jump in current results from water swelling the membrane and mechanical stress relaxation improving the membrane-electrode contact. We have seen this phenomenon in other contexts of PEM fuel cell dynamics and it is the subject on on-going investigations.

## III.5. DYNAMICS OF "STEADY STATE" OPERATION

The dynamic data presented so far were all in response to changes in system parameters. We chose to present "well-behaved" responses, so that the behavior could be easily rationalized. Dynamic data for the STR PEM fuel cell can be much more complex than what we presented here. We conclude with an example that illustrates some of the complex dynamics of the STR PEM fuel cell that are still far from being understood.

Figure 11 shows the steady state response of the STR PEM fuel cell over a 24-hour period. All the external controllable parameters were fixed. The feed flow rates were constant at 5 mL/min of $H_2$ to the anode and 10 mL/min $O_2$ to the cathode. The fuel cell temperature was fixed at 80°C and the load resistance was fixed at 20Ω. The current, voltage, and relative humidity in the effluent streams all displayed autonomous oscillations with a frequency of $10^{-4}$ Hz! The current oscillations were large amplitude, changing by a factor of 2 between 75 to 170 mA. The current oscillations overshot and undershot the plateau values at the high and low states. The effluent relative humidity at both the anode and cathode oscillated in phase with the current.

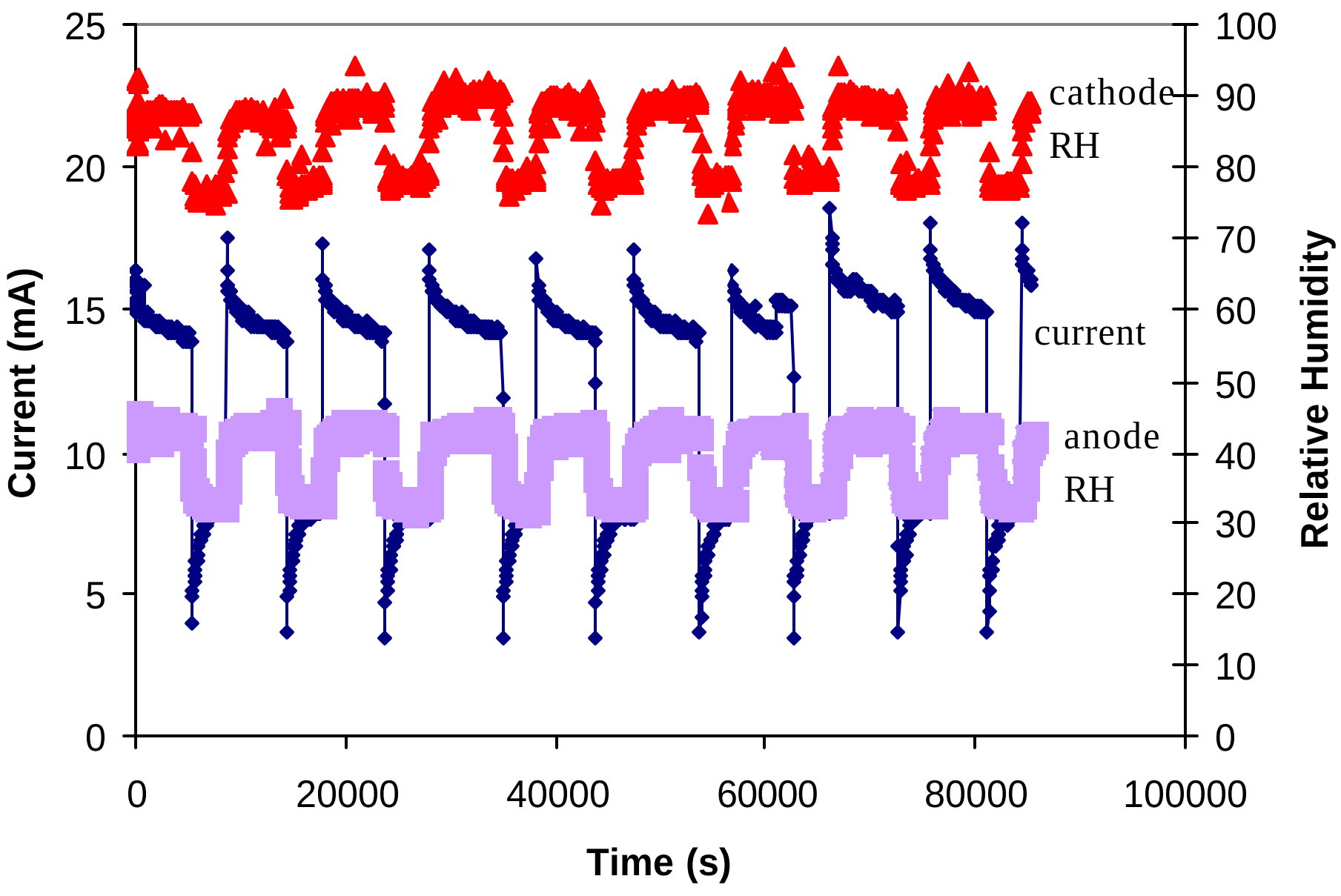

Figure 11. Steady state response of STR PEM fuel cell over a 24 hour period. The feed flow rates were 5 mL/min of $H_2$ and 10 mL/min $O_2$, the cell temperature was 80°C and the load resistance was 20 Ω.

We have observed these autonomous oscillations under a variety of conditions of temperature, flow rate and load resistance. We believe they are caused by coupling between the mechanical relaxation of the polymer membrane to changes in the water content. However, we are still a long way from understanding the physics in sufficient detail to develop predictive dynamic models for these results. Complex dynamic behavior has been anecdotally reported for fuel cell test stations, but seems to have been ignored because of lack of models that predict any such behavior. The STR PEM fuel cell displays the oscillations distinctly, and we believe by uncoupling the temporal oscillations from spatial variations we can clarify their origin and control them. Data of this quality and relative simplicity shown here is essential to understand the complex dynamics of PEM fuel cells.

## IV. CONCLUSION

Our purpose in this paper was to introduce the use of a differential reactor to study fuel cell dynamics. The data presented in this paper show that the PEM fuel cell responses are characterized by time constants varying from less than a second to thousands of seconds. The STR PEM fuel cell is a one-dimensional differential reactor that is ideally suited to examine dynamics of the coupling of reaction and transport processes in a polymer membrane fuel cell. The STR PEM has even unveiled novel behavior that suggests mechanical properties of the polymer membranes may play an important role in fuel cell dynamics.

The STR PEM fuel cell is not an optimal design of a fuel cell reactor in the sense of obtaining the highest power output or highest fuel efficiency. Its purpose is to provide a well-defined set of reactor conditions to facilitate the correlation of fuel cell operation with process parameters. We have stressed the importance of characterizing the system variables and relating them to changes in the system parameters. This approach is vital to the development of effective control systems for fuel cells.

The STR PEM fuel cell has exemplified how PEM fuel cells “ignite” and the critical role the water balance plays in the dynamics of ignition. The water activity in the membrane must equilibrate to changes in the control parameters, feed flow rates, cell temperature and load resistance. Changes in the control parameters alter the balance between water production and water removal. PEM fuel cells typically have at least two time constants associated with their transient responses. There is a very rapid response, time constant < 1 s, corresponding to the changes in external load at constant membrane water activity. There are longer responses with a time constants of ~100 s corresponding to water transport in the membrane and equilibration of the membrane water activity. Lastly we showed there are additional dynamical processes with time constants of ~1,000-10,000 s, probably due to mechanical relaxation processes that are not yet fully understood.

## V. NOMENCLATURE

$A_m$ – area of membrane
$a_w^i$ - water activity at anode (A), cathode (C), or membrane (m)
EW – equivalent weight of membrane (mass/mole of $SO_3$)
F- Faraday's constant
$\Delta G^o$ - free energy of the fuel cell reation
*i* - current
$N_w^m$ – water content in membrane (moles)
$P_i$ – partial pressure of species i (bar)
$Q_i$ – volumetric flow rate
R – gas constant
$R_m$ – effective resistance of the membrane electrode assembly
$R_L$ – external load resistance
T – fuel cell temperture
$V_{eff}$ – effective output voltage of fuel cell
$V_i$ – gas volume at anode (A) and cathode (C)
$\Delta z$ – membrane thickness
$\eta$ – overpotential of fuel cell (a function of membrane water activity and load resistance)
$\lambda_w$ – absorbed water concentration per sulfonic acid content (# water/#$SO_3$)